# W-state Analyzer and Multi-party Measurement-device-independent Quantum Key Distribution

Changhua Zhu,[1 2*] Feihu Xu,[3] & Changxing Pei[1]

[1]State Key Laboratory of Integrated Services Networks, Xidian University, Xi'an, Shaanxi 710071, China, [2]Department of Electrical & Computer Engineering, University of Toronto, Toronto, Ontario M5S 3G4, Canada, [3]Research Laboratory of Electronics, Massachusetts Institute of Technology, Cambridge, Massachusetts 02139, USA

*Corresponding author: chhzhu@xidian.edu.cn

**W-state is an important resource for many quantum information processing tasks. In this paper, we for the first time propose a multi-party measurement-device-independent quantum key distribution (MDI-QKD) protocol based on W-state. With linear optics, we design a W-state analyzer in order to distinguish the four-qubit W-state. This analyzer constructs the measurement device for four-party MDI-QKD. Moreover, we derived a complete security proof of the four-party MDI-QKD, and performed a numerical simulation to study its performance. The results show that four-party MDI-QKD is feasible over 150 km standard telecom fiber with off-the-shelf single photon detectors. This work takes an important step towards multi-party quantum communication and a quantum network.**

The quantum key distribution (QKD) protocol, which is based on the principles of quantum mechanism, is unconditionally secure in theory[1,2]. For a review, see, e.g. Ref. 3. In practice, however, a QKD system still has security loopholes due to the gap between theory and practice. Various attacks have been successfully launched through the exploration of these loopholes, e.g. a time-shift attack[4, 5], a phase-remapping attack[6], a blinding attack[7, 8], and so forth[9-11]. To close this gap, the first method is to build precise mathematical models for all the devices and refine the security proofs to include these models[12]. However, this method is challenging to implement due to the complexity of QKD components. In addition, a device-independent QKD (DI-QKD) was proposed[13, 14]. In DI-QKD, the legitimate participants during the process of communication, namely, Alice and Bob, do not need to obtain precise mathematical models for their devices, and all side-channels can be removed from QKD implementations if certain requirements can be satisfied. However, the implementation requires a loophole-free Bell test, which is still out the scope of current technology. Instead, a new protocol, measurement-device-independent QKD (MDI-QKD)[15] (for a review, see Ref. 16), was proposed. This protocol is fully practicable with current technology. Unlike security patches[17, 18], MDI-QKD can remove all detector side-channel attacks. This kind of attack is arguably the most important security loophole in conventional QKD implementations[7-11, 19]. The measurement setup in MDI-QKD can be fully untrusted and even manufactured by Eve. The experimental feasibility of MDI-QKD has been demonstrated in both the laboratory and field tests[20-23]. MDI-QKD has also attracted a lot of scientific attention from theoretical side[24-31]. In addition to the application in QKD, MDI technique can also be used in other quantum information processing tasks, such as MDI entanglement-witness[32].

In addition to the two-party QKD protocol, researchers have also proposed various multi-party QKD protocols. Generally, there are three types of multi-party QKD schemes. The first one is based on a trusted center (TC)[33], in which each user shares a secret key with the TC and builds a common session key. The second one is an entanglement-based multi-party QKD protocol. Cabello proposed a multi-party QKD protocol that uses Greenberger-Horne-Zeilinger (GHZ) states[34] and that is an extension of a two-party entanglement-based QKD protocol[2]. Chen and Lo proposed a wide class of distillation schemes for multi-party entanglement, which have been applied to implement conference key agreement[35,36]. The third one is a multi-party QKD protocol without the use of entanglement and TC. Matsumoto proposed a QKD protocol in which Alice sends the same qubits sequence to Bob and Charlie respectively, and the qubits with coincident bases are used to build a secret key after post-processing[37]. In the first type of scheme, information may be leaked since pre-shared secret bits are used repeatedly. In the second type, a perfect GHZ state should be prepared. In the third type, two prepare-and-measure QKD processes are implemented. Nevertheless, up until now, a key weakness of all multi-party quantum cryptographic protocols is the assumption that the measurement devices are trusted. As aforementioned, the occurrence of many quantum hacking attacks indicates that this is a highly unrealistic assumption.

In order to remove the demanding requirement for trusted measurement devices, we focus our attention on multi-party MDI-QKD. Appropriate entanglement states and their analyzers are the premises for the design of a multi-party MDI-QKD protocol. An elegant GHZ-type multi-party MDI-QKD protocol has been recently proposed in Ref. 38, and this protocol shows that three-party MDI-QKD is highly feasible in practice. However, Ref. 38 is primarily limited to three participants, and in a situation with more participants, the GHZ-type MDI-QKD is restricted to a very low key rate. Another potential candidate to build multi-party MDI-QKD is cluster state, but an efficient cluster-state analyzer based on linear optics remains unknown. Therefore, in a large-scale quantum Internet, a better analyzer and a different type of entanglement state are essential and required in order to design a multi-party MDI-QKD protocol and to obtain a high key rate.

$W$-state is a category of multi-particle entanglement state that can be used in a number of quantum information processing protocols[39]. $W$-state can be generated by type-II spontaneous parametric down-conversion (SPDC) and linear optical components[40,41]. In comparison with GHZ state, an important property of $W$-state is that, if one particle is traced out and projected into a specified state, the remaining particles are still entangled. That is, W-state is highly robust. Nonetheless, a $W$-state analyzer, which would enable the state of multiple particles to be projected into a $W$-state, still has to be constructed properly.

Here, we, for the first time, propose a multi-party QKD protocol based on W-state. We present the application of $W$-state in multi-party QKD, and construct a new $W$-state analyzer to distinguish the four-qubit $W$-state, based on linear optics only. With this analyzer, a four-party $W$-state MDI-QKD protocol is proposed. In this protocol, the four users, Alice, Bob, Charlie, and David, each send BB84 qubits to the central relay, Emma, with a $W$-state analyzer. The qubits with successful measurement outputs and coincident bases are used to build a secret key. The results show that the scheme is highly feasible for practically distributing the post-selected-state entanglement and for generating secure keys over a distance of more than 150 km standard telecom fiber for experimentally accessible parameter regimes. With state-of-the-art high-efficiency detectors, four-party MDI-QKD is feasible over 250 km fiber. We remark that, our protocol can be extended to the case with more participants and still remain a high key rate. All these features move an important step towards practical multi-party quantum communication.

## Results

**$W$-state and its analyzer.** In this section, a group of four-particle entanglement $W$ states is introduced, and a four-particle $W$-state analyzer based on linear optics is proposed.

$W_4$ *state*. The standard n-qubit $W$-state is defined by ($n \geq 3$)[42]

$$|W_n\rangle = 1/\sqrt{n}\left(|10\cdots0\rangle + |01\cdots0\rangle + \cdots + |00\cdots1\rangle\right) \tag{1}$$

If $n = 4$, the four-qubit $W$-state is given by

$$|W_4\rangle = 1/2\left(|1000\rangle + |0100\rangle + |0010\rangle + |0001\rangle\right) \tag{2}$$

There are nine families of states that correspond to nine different ways of entangling four qubits[43]. For $W$-state, the widely used state is the standard one, given by equation (2). Here, based on the $W_4$ state, 16 four-qubit W states can be constructed, and these states appear in Supplementary I. All these $W_4$ states form a group of orthogonal bases in a 16-dimensional Hilbert space. Any four-qubit state can be expressed as a linear combination of these 16 $W_4$ states. The protocol proposed in this paper is based on these states.

*A four-photon $W_4$ state analyzer*. The tomography of $W$-states has been a hot topic in recent years[44-46]. However, the method for designing an analyzer to verify a $W$-state is still an open question.

In fact, a four-qubit $W_4$ state can be expressed by Bell states, which is presented as below.

$$|W_{4,0}\rangle = 1/2\left[\left(|\phi^+\rangle_{12} + |\phi^-\rangle_{12}\right)|\psi^+\rangle_{34} + |\psi^+\rangle_{12}\left(|\phi^+\rangle_{34} + |\phi^-\rangle_{34}\right)\right], \tag{3}$$

where $|\phi^+\rangle$, $|\phi^-\rangle$, and $|\psi^+\rangle$ are three Bell states. From equation (3), we find that it is possible to design a $W_4$ state analyzer based on a Bell-state analyzer. Indeed, this is our method to construct the $W_4$ state analyzer.

Generally, with an optimal linear optics-based scheme and without the use of auxiliary photons, only two out of four Bell states can be distinguished[47]. However, an important time-bin-based Bell-state analyzer can distinguish three out of four Bell states[48]. Its schematic representation is shown in Fig.1. In this scheme, the qubit is encoded with time bins[49]. The qubit $|0\rangle$ ($|1\rangle$) corresponds to a photon in state $\hat{a}_{t_0}^\dagger|0\rangle$ ($\hat{a}_{t_1}^\dagger|0\rangle$) under Z-basis, or in state $1/2(\hat{a}_{t_0}^\dagger + \hat{a}_{t_1}^\dagger)|0\rangle$ ($1/2(\hat{a}_{t_0}^\dagger - \hat{a}_{t_1}^\dagger)|0\rangle$) under X-basis, where $t_1 = t_0 + \tau$ and $\tau$ is a constant time. The device consists of two beam splitters, $BS_1$ and $BS_2$, two fibers with time delay $\tau$, and two single photon detectors, $D_1$ and $D_2$, all of which build a time-bin interferometer.

In Fig.1, let $\hat{a}_{p,t}^\dagger$ denote the creation operators in spatial mode $p$ ($p = a,b,c,d,e,f$) and temporal mode $t$ ($t = t_0$ or $t_1$ for modes $a,b,c,d$; $t = t_0, t_1$ or $t_2$ for modes $e$ and $f$). After passing through $BS_1$, let the relative phase between the transmitted light field and the reflected light field be $\pi$., then the operators evolve as follows[50]

$$\hat{a}^\dagger_{a,t} \to 1/\sqrt{2}\left(-i\hat{a}^\dagger_{c,t} + \hat{a}^\dagger_{d,t}\right) , \tag{4}$$

$$\hat{a}^\dagger_{b,t} \to 1/\sqrt{2}\left(\hat{a}^\dagger_{c,t} - i\hat{a}^\dagger_{d,t}\right) . \tag{5}$$

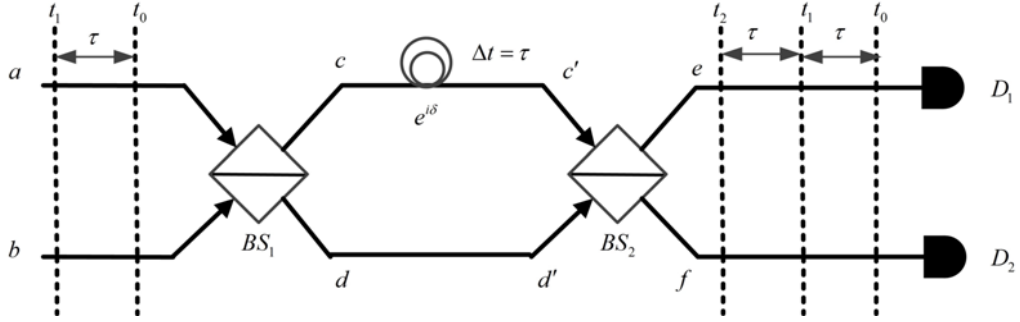

**Figure 1. A schematic diagram of Bell-state measurement[48].** $BS_1$ and $BS_2$ are ideal 50/50 optical beam splitters that have equal reflection and transmission coefficients and no absorption loss. The delay $\Delta t$ derived from the path length difference of the interferometer equals $\tau$. When two qubits enter the interferometer, the output state is a mixture of photons in two spatial modes ($e$ and $f$) and three temporal modes ($t_0, t_1$ and $t_2$). Three Bell states can be distinguished through an analysis of different combinations of these modes of four photons.

Next, after the time-bin interferometer, the creation operators evolve into

$$\hat{a}^\dagger_{a,t} \to 1/2\left(-\hat{a}^\dagger_{e,t} + e^{i\delta}\hat{a}^\dagger_{e,t+\tau} + i\hat{a}^\dagger_{f,t} + ie^{i\delta}\hat{a}^\dagger_{f,t+\tau}\right) , \tag{6}$$

$$\hat{a}^\dagger_{b,t} \to 1/2\left(\hat{a}^\dagger_{f,t} - e^{i\delta}\hat{a}^\dagger_{f,t+\tau} + i\hat{a}^\dagger_{e,t} + ie^{i\delta}\hat{a}^\dagger_{e,t+\tau}\right) , \tag{7}$$

where $\delta$ is the phase derived from the path length difference in the interferometer[48]. Equations (6) and (7) indicate that the photons may arrive at $D_1$ or $D_2$ at different time instants, $t_0$, $t_1$, or $t_2$, according to input states, as shown in Fig.1. From the output coincidence, in principle, Bell-states $\left|\psi^+\right\rangle_{ab}$ can be detected with 100% probability, and $\left|\psi^-\right\rangle_{ab}$ and $\left|\phi^+\right\rangle_{ab}$ can be detected with 50% probability, respectively[48].

The W-state analyzer shown in Fig. 2 is proposed. The qubits $\left|0\right\rangle$ and $\left|1\right\rangle$ are also encoded with the time-bin. At the first stage, the states of the photons in spatial modes $a$ and $b$ evolve into the states at modes $e$ and $f$, according to equations (6) and (7). In ways that are also similar to equations (6) and (7), the states of the photons in spatial modes $c$ and $d$ evolve into the ones in modes $g$ and $h$; the ones in modes $f$ and $g$ evolve into the ones in modes $j$ and $k$; the ones in modes $e$ and $h$ evolve into the ones in modes $l$ and $m$.

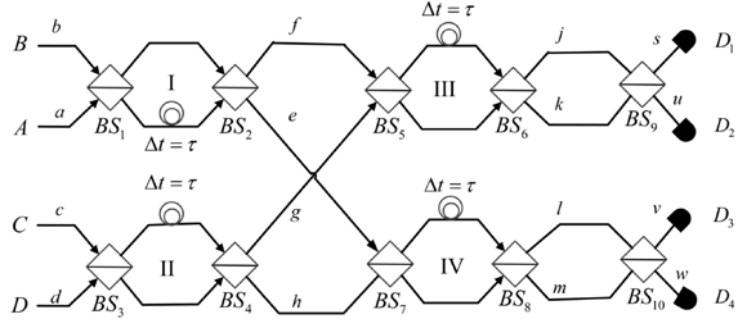

**Figure 2. A schematic diagram of a $W$-state analyzer.** The analyzer consists of four time-bin interferometers. Each interferometer is the same as the one shown in Fig.1. Two photons from port A and B enter into the interferometer I. The photon in spatial mode $e$ enters into interferometer IV, and the photon in spatial mode $f$ enters into interferometer III. Another two photons from port C and D enter into interferometer II. The photon in spatial mode $g$ enters into interferometer III, and the photon in spatial mode $h$ enters into interferometer IV. The state of photons at the output is in a superposition state of four spatial modes ($s, u, v, w$) and four temporal modes ($t_0, t_1, t_2$ and $t_3$). By analysis of different combinations of these modes, four out of sixteen $W$-states can be distinguished.

After $BS_9$ and $BS_{10}$, the states of photons evolve into

$$\hat{a}_{j,t}^\dagger \to 1/\sqrt{2}\left(-i\hat{a}_{s,t}^\dagger + \hat{a}_{u,t}^\dagger\right) \tag{8}$$

$$\hat{a}_{k,t}^\dagger \to 1/\sqrt{2}\left(\hat{a}_{s,t}^\dagger - i\hat{a}_{u,t}^\dagger\right) \tag{9}$$

$$\hat{a}_{l,t}^\dagger \to 1/\sqrt{2}\left(-i\hat{a}_{v,t}^\dagger + \hat{a}_{w,t}^\dagger\right) \tag{10}$$

$$\hat{a}_{m,t}^\dagger \to 1/\sqrt{2}\left(\hat{a}_{v,t}^\dagger - i\hat{a}_{w,t}^\dagger\right) \tag{11}$$

Based on equations (6) - (11), the output detection modes of each of 16 $W_4$ states can be obtained. As an example, the state $|W_{4,0}\rangle$ is discussed. Its operator form is

$$|W_{4,0}\rangle = 1/2(|0001\rangle + |0010\rangle + |0100\rangle + |1000\rangle)$$

$$= 1/2\left[\hat{a}_{a,t_0}^\dagger \hat{a}_{b,t_0}^\dagger \hat{a}_{c,t_0}^\dagger \hat{a}_{d,t_1}^\dagger + \hat{a}_{a,t_0}^\dagger \hat{a}_{b,t_0}^\dagger \hat{a}_{c,t_1}^\dagger \hat{a}_{d,t_0}^\dagger + \hat{a}_{a,t_0}^\dagger \hat{a}_{b,t_1}^\dagger \hat{a}_{c,t_0}^\dagger \hat{a}_{d,t_0}^\dagger + \hat{a}_{a,t_1}^\dagger \hat{a}_{b,t_0}^\dagger \hat{a}_{c,t_0}^\dagger \hat{a}_{d,t_0}^\dagger\right]|0\rangle \tag{12}$$

Then, by using equations (6) - (11), $|W_{4,0}\rangle$ evolves into

$$|W_{4,0}\rangle \to 1/2048\left\{64e^{i2\delta}\hat{a}_{s,t_0}^\dagger\left(\hat{a}_{s,t_1}^\dagger\right)^3 - 192e^{i6\delta}\hat{a}_{s,t_1}^\dagger \hat{a}_{s,t_2}^\dagger \left(\hat{a}_{w,t_2}^\dagger\right)^2 - 64e^{i3\delta}\hat{a}_{s,t_1}^\dagger \left(\hat{a}_{u,t_1}^\dagger\right)^2 \hat{a}_{w,t_1}^\dagger + \cdots + \right.$$
$$128e^{i2\delta}\hat{a}_{s,t_0}^\dagger \hat{a}_{u,t_1}^\dagger \hat{a}_{v,t_0}^\dagger \hat{a}_{w,t_1}^\dagger + 128e^{i2\delta}\hat{a}_{s,t_0}^\dagger \hat{a}_{u,t_1}^\dagger \hat{a}_{v,t_1}^\dagger \hat{a}_{w,t_1}^\dagger + 128e^{i2\delta}\hat{a}_{s,t_0}^\dagger \hat{a}_{u,t_2}^\dagger \hat{a}_{v,t_0}^\dagger \hat{a}_{w,t_1}^\dagger + \cdots + \tag{13}$$
$$\left. 128e^{i4\delta}\hat{a}_{v,t_0}^\dagger \hat{a}_{v,t_1}^\dagger \hat{a}_{w,t_1}^\dagger \hat{a}_{w,t_3}^\dagger + 128e^{i4\delta}\hat{a}_{v,t_0}^\dagger \hat{a}_{v,t_2}^\dagger \hat{a}_{w,t_1}^\dagger \hat{a}_{w,t_2}^\dagger \right\}|0\rangle$$

There are 200 terms in equation (13). That means that the output state is a superposition of 200 states. Each of the states is called a detection mode. Here, detection mode means SPD clicks at some spatial modes and time-bins; e.g., $\hat{a}_{s,t_i}^\dagger \hat{a}_{u,t_j}^\dagger \hat{a}_{v,t_k}^\dagger \hat{a}_{w,t_l}^\dagger$ means that photons clicks occur in spatial modes $s$, $u$, $v$, and $w$ at the time instant $t_i$, $t_j$, $t_k$, and $t_l$ ($i,j,k,l = 0,1,2,3$), respectively. $\left(\hat{a}_{s,t_i}^\dagger\right)^2 \hat{a}_{u,t_j}^\dagger \hat{a}_{v,t_k}^\dagger$ means that two photons arrive in $s$ mode at $t_i$. One photon occurs in $u$ mode at $t_j$, and one photon occurs in $v$ mode at time $t_k$, respectively. The spatial and temporal modes are shown in Fig. 3.

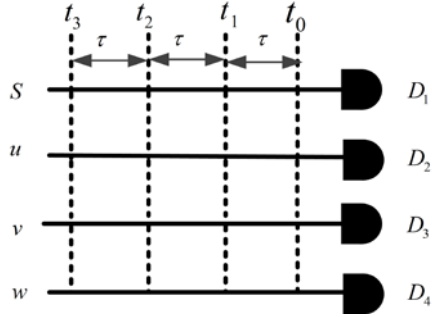

**Figure 3. The spatial and temporal modes at the output of the $W$-state analyzer.** There are four spatial modes (i.e., $s, u, v$, and $w$) and four temporal modes (i.e., $t_0, t_1, t_2$, and $t_3$). The time separation between time-bins is $\tau$. The output state is a superposition of spatial modes and temporal modes.

All detection modes of 16 $W_4$ states have been obtained. By comparing the detection modes among different states, the unique modes belonging to one specific state are obtained. These modes make one state distinguishable from the others. Here, only the modes in which the SPD click derives from one photon are taken into account; i.e., all four SPDs in each mode have a click. There are four $W_4$ states, i.e., $|W_{4,0}\rangle$, $|W_{4,1}\rangle$, $|W_{4,c}\rangle$, and $|W_{4,d}\rangle$, that can be identified with the proposed analyzer. Their detection modes are shown in Table 1. The success rate is determined by the corresponding coefficients of output states. For states $|W_{4,0}\rangle$ and $|W_{4,c}\rangle$, the probability of successful detection is $D_{p0} = [128/2048]^2 \times 12 = 0.0469$. For states $|W_{4,1}\rangle$ and $|W_{4,d}\rangle$, the probability of successful detection is $D_{p1} = [128/2048]^2 \times 4 = 0.0156$. Therefor the total success probability is $D_p = 1/16(2 \times D_{p0} + 2D_{p1}) = 0.78\%$. These four states can be applied to build keys among four users, a process that will be discussed in section III.

It is worth mentioning that four other states, i.e., $|W_{4,2}\rangle$, $|W_{4,3}\rangle$, $|W_{4,e}\rangle$ and $|W_{4,f}\rangle$, can also be distinguished if photon-number-resolving detectors can be used. In addition, the detection probabilities of states $|W_{4,0}\rangle$, $|W_{4,1}\rangle$, $|W_{4,c}\rangle$, and $|W_{4,d}\rangle$ can also be increased with this type of detector.

**TABLE 1.** Distinguishable $W_4$ states and their detection modes. Four $W_4$ states, i.e., $|W_{4,0}\rangle$, $|W_{4,1}\rangle$, $|W_{4,c}\rangle$, and $|W_{4,d}\rangle$, can be identified by the proposed analyzer. Detection modes $s_i u_j v_k w_l$ mean that photons clicks occur in the spatial modes $s$, $u$, $v$, and $w$, and at the temporal modes $t_i$, $t_j$, $t_k$, and $t_l$ ($i, j, k, l = 0, 1, 2, 3$), respectively.

| No. | Distinguished states | Detection modes | Success probability |
|---|---|---|---|
| I | $|W_{4,0}\rangle$ | $s_0 u_1 v_0 w_2$, $s_0 u_1 v_1 w_1$, $s_0 u_1 v_1 w_3$, $s_0 u_1 v_2 w_2$, $s_0 u_2 v_0 w_1$, $s_0 u_2 v_0 w_3$, $s_0 u_3 v_0 w_2$, $s_1 u_1 v_0 w_1$, $s_1 u_1 v_2 w_1$, $s_1 u_3 v_0 w_1$, $s_2 u_1 v_1 w_1$, $s_2 u_2 v_0 w_1$ | 0.0469 |
| II | $|W_{4,1}\rangle$ | $s_0 u_1 v_0 w_3$, $s_0 u_1 v_2 w_1$, $s_0 u_3 v_0 w_1$, $s_2 u_1 v_0 w_1$ | 0.0156 |
| III | $|W_{4,c}\rangle$ | $s_0 u_2 v_2 w_3$, $s_0 u_3 v_1 w_3$, $s_1 u_1 v_2 w_3$, $s_1 u_3 v_0 w_3$, $s_1 u_3 v_2 w_3$, $s_2 u_1 v_2 w_2$, $s_2 u_2 v_2 w_1$, $s_2 u_2 v_2 w_3$, $s_2 u_3 v_0 w_2$, $s_2 u_3 v_1 w_1$, $s_2 u_3 v_1 w_3$, $s_2 u_3 v_2 w_2$ | 0.0469 |
| IV | $|W_{4,d}\rangle$ | $s_0 u_3 v_2 w_3$, $s_2 u_1 v_2 w_3$, $s_2 u_3 v_0 w_3$, $s_2 u_3 v_2 w_1$ | 0.0156 |

**Measurement-device-independent quantum key distribution based on W-state.** In this section, we propose a four-party MDI-QKD protocol based on $W_4$ state and the analyzer presented in the previous section. The security of the protocol is also proved.

*The protocol.* Conceptually, the four-party MDI-QKD can be implemented based on a time-reversal $W_4$ state protocol. In this protocol, each of the four users can prepare an entangled EPR photon pairs, keep one photon from each pair, and send the other photon to the central relay. Then projective measurement on the state of the photons can be performed by the relay. If the state is projected into a $W_4$ state by the relay, the state of the remaining four photons in the users is projected to the same $W_4$ state. Through the use of the idea of a virtual qubit[16], a four-party MDI-QKD scheme can be constructed.

The proposed setup of four-party MDI-QKD protocol is shown in Fig. 4. There are four participants, i.e., Alice, Bob, Charlie, and David. Photons from single photon sources (SPS) are encoded with time-bin.

Generally, weak coherent pulse (WCP) sources combined with decoy state technology[51-53] can also be used to replace the SPS. Here, SPS is used to simplify the discussion.

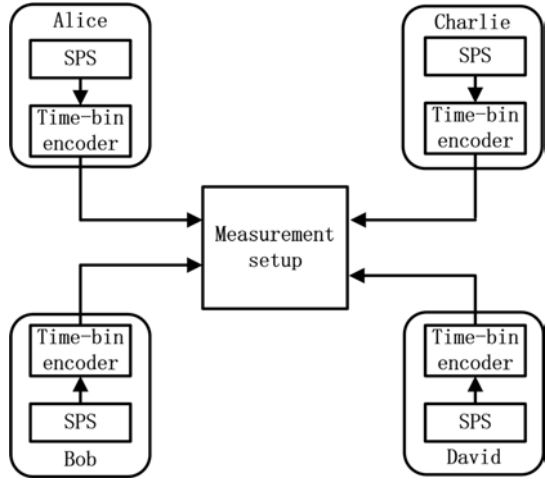

**Figure 4. Basic setup of a $W_4$-based four-party MDI-QKD protocol**. Alice, Bob, Charlie, and David prepare single photon pulses in a different BB84 time-bin coded state, which is selected independently and at random for each signal. The time-bin encoder can follow the design proposed in Ref. 20. Inside the measurement setup, signals from Alice, Bob, Charlie, and David are sent into a W-state analyzer (see Fig. 2). Successful output corresponds to the observation of one of four $W_4$ states shown in Table 1. The four participants' laboratories are well shielded from the eavesdropper, and the measurement setup can be untrusted.

The procedures of the protocol are as follows:

(1) **Preparing**: Each one of the participants, i.e., Alice, Bob, Charlie, and David, prepares single photons, which are in the four possible BB84 time-bin states (i.e., $|0\rangle$, $|1\rangle$, $|+\rangle$, and $|-\rangle$) and sends them to an untrusted relay, Emma, with an analyzer in the middle. The preparation processes are implemented by single photon sources and a time-bin encoder.

(2) **Measuring**: Emma performs $W_4$ state measurement by using the analyzer in Fig. 2. Then the incoming signals are projected into a $W_4$ state.

(3) **Sifting**: Emma uses public channels to announce the events in which she obtained successful outputs; i.e., some of the states in Table 1 are identified. When all participants use the rectilinear (Z) basis, two of them announce their bits, and the other two perform operations according to the scenarios shown in Table 2.

In addition to the case that all participants encode their qubits in Z basis, another case is that they encode their qubits in X basis. For the latter, the W-states can be described as states $|+\rangle$ and $|-\rangle$, e.g.,

$$|W_{4,0}\rangle = 1/4 \big[ |++\rangle(2|++\rangle + |+-\rangle + |-+\rangle) + (|+-\rangle + |-+\rangle)(|++\rangle - |--\rangle) - |--\rangle(|+-\rangle + |-+\rangle + 2|--\rangle) \big] \tag{14}$$

**TABLE 2.** Four participants' post-selection after Emma announces a successful output of states $|W_{4,0}\rangle$ or $|W_{4,1}\rangle$ ($|W_{4,c}\rangle$ or $|W_{4,d}\rangle$). Any two participants announce their classic bits. If the bits are "00" ("11"), the other two participants can obtain the raw key bits; i.e., one of them flips his or her bits. For example, when Alice and Bob announce classic bits "00" ("11"), one of the pair Charlie and David flips his bits. This way, any two participants can perform QKD. The optical quantum channel need not be changed.

| Announced bits | | | | Participants who obtain the key bits and their operation |
|---|---|---|---|---|
| Alice | Bob | Charlie | David | |
| 0 (1) | 0 (1) | - | - | Charlie & David, one of their bits flips. |
| 0 (1) | - | 0 (1) | - | Bob & David, one of their bits flips. |
| 0 (1) | - | - | 0 (1) | Bob & Charlie, one of their bits flips. |
| - | 0 (1) | 0 (1) | - | Alice & David, one of their bits flips |
| - | 0 (1) | - | 0 (1) | Alice & Charlie, one of their bits flips. |
| - | - | 0 (1) | 0 (1) | Alice & Bob, one of their bits flips. |

and

$$|W_{4,c}\rangle = 1/4 \big[|++\rangle(2|++\rangle - |+-\rangle - |-+\rangle) - (|+-\rangle + |-+\rangle)(|++\rangle - |--\rangle) + |--\rangle(|+-\rangle + |-+\rangle - 2|--\rangle)\big] \quad (15)$$

In this case, the first two announce the values of the qubits ($|+-\rangle$ or $|-+\rangle$), and the other two perform phase error rate estimation.

(4) **Post-processing**: After obtaining the sifted key, the two participants perform information reconciliation and privacy amplification. The suggestion is that an error correction code-based reconciliation protocol be used, since the interactive protocol, e.g., Cascade[54], requires many communications. A low-density parity-check (LDPC) code-based reconciliation scheme[55] can be used.

*Security analysis.* The security of the four-party $W$-state-based MDI-QKD protocol is inspired by the security of a time reversed W-state-based QKD protocol.

First, we briefly introduce the $W$-state-based QKD protocol. In a three-party W-state-based QKD protocol, three particles in $W_3$ state are distributed to three participants respectively. The announcement of the measurement bases and the measurement results of one participant enables the other two to perform key distribution or security verification. The protocol can be extended to the one with four participants. Compared with a three-party QKD protocol, in the one with four participants, two participants announce their measurement bases and results, and the other two are in a maximally entangled Bell state and can obtain a secret key.

Secondly, it can be demonstrated that a time reversed $W$-state-based QKD protocol exists as the same as the time reversed EPR protocol[56]. With reference to the two-party MDI-QKD protocol[15], the idea of a virtual qubit is also used. One can imagine that each of four participants prepares an EPR entanglement state, sends one qubit to Emma, and retains the other qubit as a virtual qubit. The virtual qubit is subsequently measured, and a BB84 state is thus prepared. In principle, each one could keep his or her virtual qubit in his or her memory and delay his or her measurement of it. Only after Emma has announced that she has obtained a successful outcome will each perform a measurement on his or her virtual qubit in order to decide which state he or she is sending to Emma. Furthermore, it is shown that W-state can be prepared among four participants by entanglement swapping, while each participant prepares an EPR pair initially. So, in such a virtual qubit setting, the protocol is equivalent to an entanglement-based protocol. Alice, Bob, Charlie, and David share quadruple qubits in their quantum memories, and they can compute the quantum bit error rate (QBER) on their virtual qubits on a special basis.

**Key rate of the four-party MDI-QKD protocol.** The key rate of the W-state-based MDI-QKD protocol is evaluated with SPS. According to the procedures described in section III.A, any two participants can build a secret key after Emma announces successful outputs, and the other two participants' classic bits are 00 or 11. So the key rate can be obtained by referring to the case of two-party MDI-QKD[15] and to the basic work of Shor and Preskill[57]. The difference between the four-party and the two-party MDI-QKD is that the gain in the four-party one refers to the joint probability that Emma announces successful output and two of participants' classic bits are 00 (or 11), according to Table 2. Since any two participants can build a secret key, the maximum information loss value in data reconciliation and the privacy amplification processes of each pair are considered. So the key rate can be given as

$$R_0 = qQ_1 \left\{ 1 - Max\left[ H_2\left(e_{cd}^X\right), H_2\left(e_{bd}^X\right), H_2\left(e_{bc}^X\right), H_2\left(e_{ad}^X\right), H_2\left(e_{ac}^X\right), \right.\right.$$
$$\left. H_2\left(e_{ab}^X\right) \right] - Max\left[ H_2\left(e_{cd}^Z\right), H_2\left(e_{bd}^Z\right), H_2\left(e_{bc}^Z\right), H_2\left(e_{ad}^Z\right), \right.$$
$$\left.\left. H_2\left(e_{ac}^Z\right), H_2\left(e_{ab}^Z\right) \right] \right\} \tag{16}$$

where $e_{jk}^X$ ($e_{jk}^Z$) denotes the QBER between participants $j$ and $k$ under X (Z) basis, given that each of Alice ($a$), Bob ($b$), Charlie ($c$), and David ($d$) sends single photon states, $j,k = a,b,c,d$; $Q_1$ denotes the gain (the joint probability of Emma's announcement of a successful detection in the Z basis, and also of the announced classic bits being 00 or 11, according to Table 2). $q$ means the basis reconciliation factor; $H_2(x)$ is the binary entropy function with parameter $x$ given by $H_2(x) = -x\log_2(x) - (1-x)\log_2(1-x)$.

In this protocol, the QBER in Z basis equals the one in X-basis under SPS, i.e. $e_{jk}^X = e_{jk}^Z$. The assumption is that there is no misalignment error, that the data size is infinite, and that the ideal reconciliation algorithm is applied. There is also an assumption that the quantum channels between the participants and Emma are identical. For the sake of simplicity, if we assume $e_1$ to be $e_{jk}^Z$, then equation (16) can be reduced to

$$R_0 = qQ_1 \left\{ 1 - 2H_2(e_1) \right\} \tag{17}$$

Let the probability of Z basis be nearly one, i.e., $q \approx 1$. $Q_1$ can be estimated as (the detailed for obtaining $Q_1$ and $e_1$ are shown in Methods and Supplementary III)

$$Q_1 = (1-Y_0)^{12}\left[1024(1-\eta)^4 Y_0^4 + 1440\eta(1-\eta)^3 Y_0^3 + 496\eta^2(1-\eta)^2 Y_0^2 + 49\eta^3(1-\eta)Y_0 + 8(D_{p0}+D_{p1})\eta^4\right]/128 \qquad (18)$$

and the QBER $e_1$ as

$$e_1 = \frac{1}{16Q_1}(1-Y_0)^{12}\left[64(1-\eta)^4 Y_0^4 + 90\eta(1-\eta)^3 Y_0^3 + 31\eta^2(1-\eta)^2 Y_0^2 + 3\eta^3(1-\eta)Y_0\right] \qquad (19)$$

where $\eta$ is the channel transmittance between the participant and the analyzer, $\eta = 10^{-\alpha \cdot l/10} \cdot \eta_d$; $\alpha$ and $l$ are the attenuation coefficient and fiber length between the participant and analyzer, and $\eta_d$ is the detection efficiency of a single photon detector. Here, it is assumed that each SPD at each time instant has the same detection efficiency.

In numerical simulation, the parameters include the detection efficiency $\eta_d$, the background count rate $Y_0$ and the attenuation coefficient $\alpha$. Let $\eta_d$ be 14.5% and $Y_0$ be $6.02\times 10^{-6}$. These values are chosen from the 144-Km QKD experiment reported in Ref. 58. A superconducting nanowire single-photon detector (SNSPD) with a detection efficiency of 93%, as reported by Marsili et.al.[59], is also used. Parameter $\alpha$ is set by a typical value, 0.2. The simulation results of the asymptotic key rate are shown in Fig. 5. The secure transmission distance between two participants is about 180 km for a detection efficiency of 14.5%, and is about 260 km for a detection efficiency of 93%. The distance is 100 km and 180 km for two detectors when the key rate is about $10^{-10}$.

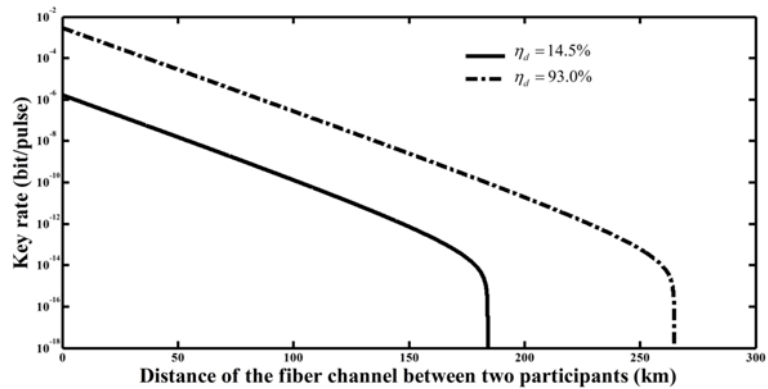

**Figure 5 Key rates with different detection efficiencies.** Both curves are key rates with single photon sources (SPSs). The solid curve is the one with a detection efficiency of 14.5%. The dash-and-dot is the one with the higher detection efficiency of 93%.

## Discussion

In practice, the SPS may still be challenging with current technology. However, based on the so-called decoy state method[51-53], one can simply replace the SPS with weak coherent pulses (WCP) or parametric down-conversion (PDC) sources. As noted already in Ref. 38 regarding the three-party MDI-QKD, the decoy state analysis and the finite-key analysis are similar to the initial two-party MDI-QKD protocol[24-28]. Therefore, the expectation is that, with decoy states, the results here can be easily extended to the cases with WCP and PDC sources.

In our proposal, any two of four parties can share a secure key bit. This is compatible to the usual network scenario, in which any two parties in the network can perform secure communications. There are several advantages as compared to the initial two-party MDI-QKD protocol. First, our proposal is faster in sharing key bits when the parties are reassigned. This is because the quantum channel is not required to be initialized. Second, the group key can also be built if one party serves as a controller. Finally, the initial MDI-QKD requires a clever design of fast and low-loss optical switches for a network setting, which might be challenging in a large-scale network. In contrast, our scheme does not have such requirement.

In the conclusion, we proposed a four-party W-state-based MDI-QKD protocol, in which any two of four participants can build secret keys, when the W-state analyzer announces a successful output, and the other two participants' classic bits sent are 00 (the distinguished states are $|W_{4,0}\rangle$ or $|W_{4,1}\rangle$) or 11 (the distinguished states are $|W_{4,c}\rangle$ or $|W_{4,d}\rangle$). Since the time-bin coded MDI-QKD protocol was verified to be feasible[20, 22, 29], and several schemes of SPS (e.g. quantum dot SPS[60]) have been presented, the proposed W-state analyzer can be implemented with current technology. The work presented here puts forward an important avenue for practical multi-party quantum communication.

**Methods**

**W-state preparation based on entanglement swapping.** A process of entanglement swapping for generating $W_4$-state is shown in Fig.6.

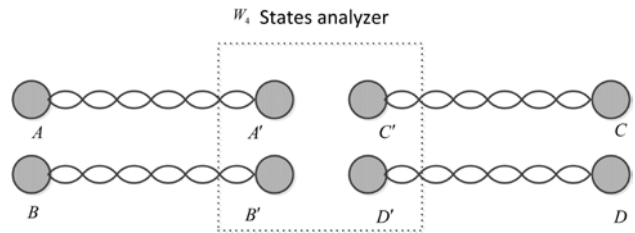

**Figure 6. A schematic diagram of entanglement swapping for generating $W_4$ state.** Each of Alice, Bob, Charlie and David prepares an EPR pair and sends half of them to Emma with a $W_4$ State analyzer.

In Fig.6, all the photon pairs $A$ and $A'$, $B$ and $B'$, $C$ and $C'$, $D$ and $D'$ are in Einstein-Podolsky-Rosen (EPR) entangled states. When the state of 4 photons, $A', B', C'$ and $D'$, are projected into to any $W_4$ state, the state of remaining four photons, $A, B, C$ and $D$, is projected into the corresponding $W_4$ sate. The detailed processes are shown in Supplementary II.

**Estimation of $Q_1$ and $e_1$.** We assume that there is no misalignment error (i.e. all mismatches in quantum channels are perfectly compensated), the four optical channels are identical, and ideal single photon sources are used. Therefore, the qubit error derives from dark counts of SPDs. As mentioned in Results, we only take into account the case in which SPD count in each spatial-temporal mode derives from no more than one photon.

For Emma, a successful output can be obtained from five cases: (1) all four counts of SPDs derive from background noise (dark counts); (2) one count derives from photon detection and the other three counts derive from background noise; (3) two counts derive from photon detections and the other two counts derive from background noise; (4) three counts derive from photon detections and the other one derives from background noise; (5) all four counts derive from photons. We assume that Alice (a) and Bob (b) announce their classical bits, while, Charlie (c) and David (d) try to generate secret key bits. In cases (2) - (5) we compute the output states of photons successfully passing through the analyzer and their probabilities. Then, we obtain the gain at each case according to the modes in Table 1. The details for obtaining $Q_1$ and $e_1$ are shown in Supplementary III.

**Acknowledgments**

The authors would like to thank Zhiyuan Tang and Kejin Wei for helpful discussions, and thank Hoi-Kwong Lo for his inspiring proposal and brilliant comments. The authors also acknowledge the financial support from the Office of Naval Research (ONR), the Air Force Office of Scientific Research (AFOSR), the National Natural Science Foundation of China No. 61372076 and No. 61301171, the 111 Project (No. B08038), and China Scholarship Council (No. 201308615037). This work was done when the first author was a visiting scholar at University of Toronto.

**Author Contributions**

C.Z. and F.X. designed the new protocol, did the security analysis and key rate calculation. C. P. supervised the project. All authors wrote and reviewed the manuscript.

**Additional information**

Competing financial interests: The authors declare no competing financial interests.

# Supplementary materials: W-state Analyzer and Multi-party Measurement-device-independent Quantum Key Distribution

Changhua Zhu[1,2*], Feihu Xu[3], & Changxing Pei[1]

[1]State Key Laboratory of Integrated Services Networks, Xidian University, Xi'an, Shaanxi 710071, China

[2]Department of Electrical & Computer Engineering, University of Toronto, Toronto, Ontario M5S 3G4, Canada

[3]Research Laboratory of Electronics, Massachusetts Institute of Technology, Cambridge, Massachusetts 02139, USA

*Corresponding author: chhzhu@xidian.edu.cn

## I. 16 $W_4$ states

The 16 $W_4$ states are given as follows:

$$|W_{4,0}\rangle = 1/2(|0001\rangle + |0010\rangle + |0100\rangle + |1000\rangle) \tag{S1}$$

$$|W_{4,1}\rangle = 1/2(|0001\rangle - |0010\rangle - |0100\rangle + |1000\rangle) \tag{S2}$$

$$|W_{4,2}\rangle = 1/2(|0001\rangle - |0010\rangle + |0100\rangle - |1000\rangle) \tag{S3}$$

$$|W_{4,3}\rangle = 1/2(|0001\rangle + |0010\rangle - |0100\rangle - |1000\rangle) \tag{S4}$$

$$|W_{4,4}\rangle = 1/2(|0000\rangle + |1100\rangle + |1010\rangle + |1001\rangle) \tag{S5}$$

$$|W_{4,5}\rangle = 1/2(|0000\rangle - |1100\rangle - |1010\rangle + |1001\rangle) \tag{S6}$$

$$|W_{4,6}\rangle = 1/2(|0000\rangle - |1100\rangle + |1010\rangle - |1001\rangle) \tag{S7}$$

$$|W_{4,7}\rangle = 1/2(|0000\rangle + |1100\rangle - |1010\rangle - |1001\rangle) \tag{S8}$$

$$|W_{4,8}\rangle = 1/2(|0011\rangle + |0101\rangle + |0110\rangle + |1111\rangle) \tag{S9}$$

$$|W_{4,9}\rangle = 1/2(|0011\rangle - |0101\rangle - |0110\rangle + |1111\rangle) \tag{S10}$$

$$|W_{4,a}\rangle = 1/2(|0011\rangle - |0101\rangle + |0110\rangle - |1111\rangle) \tag{S11}$$

$$|W_{4,b}\rangle = 1/2(|0011\rangle + |0101\rangle - |0110\rangle - |1111\rangle) \tag{S12}$$

$$|W_{4,c}\rangle = 1/2(|0111\rangle + |1011\rangle + |1101\rangle + |1110\rangle) \tag{S13}$$

$$|W_{4,d}\rangle = 1/2(|0111\rangle - |1011\rangle - |1101\rangle + |1110\rangle) \tag{S14}$$

$$|W_{4,e}\rangle = 1/2(|0111\rangle - |1011\rangle + |1101\rangle - |1110\rangle) \tag{S15}$$

$$|W_{4,f}\rangle = 1/2(|0111\rangle + |1011\rangle - |1101\rangle - |1110\rangle) \tag{S16}$$

The state $|W_{4,0}\rangle$ given by equation (S1) is a standard $W_4$ state and belongs to one family states defined in Ref. 1. The states $|W_{4,1}\rangle$, $|W_{4,2}\rangle$ and $|W_{4,3}\rangle$ can be constructed by performing unitary operations *IZZI*, *ZIZI* and *ZZII* on state $|W_{4,0}\rangle$, respectively, where $I$ is the unit operator and $Z$ is the Pauli $Z$ operator. The states $|W_{4,c}\rangle$, $|W_{4,d}\rangle$, $|W_{4,e}\rangle$ and $|W_{4,f}\rangle$ can be constructed by performing *XXXX* operations on the states $|W_{4,0}\rangle$, $|W_{4,1}\rangle$, $|W_{4,2}\rangle$ and $|W_{4,3}\rangle$, respectively, where $X$ is the Pauli $X$ operator. The state $|W_{4,4}\rangle$ can be obtained by performing unitary operation U

$$U = \begin{bmatrix} 0 & 1 & 0 & 0 & 0 & 0 & 0 & 0 & 0 & 0 & 0 & 0 & 0 & 0 & 0 & 0 \\ 1 & 0 & 0 & 0 & 0 & 0 & 0 & 0 & 0 & 0 & 0 & 0 & 0 & 0 & 0 & 0 \\ 0 & 0 & 0 & 1 & 0 & 0 & 0 & 0 & 0 & 0 & 0 & 0 & 0 & 0 & 0 & 0 \\ 0 & 0 & 0 & 0 & 0 & 1 & 0 & 0 & 0 & 0 & 0 & 0 & 0 & 0 & 0 & 0 \\ 0 & 0 & 0 & 0 & 0 & 0 & 1 & 0 & 0 & 0 & 0 & 0 & 0 & 0 & 0 & 0 \\ 0 & 0 & 0 & 0 & 0 & 0 & 0 & 1 & 0 & 0 & 0 & 0 & 0 & 0 & 0 & 0 \\ 0 & 0 & 0 & 0 & 0 & 0 & 0 & 0 & 0 & 1 & 0 & 0 & 0 & 0 & 0 & 0 \\ 0 & 0 & 0 & 0 & 0 & 0 & 0 & 0 & 0 & 0 & 1 & 0 & 0 & 0 & 0 & 0 \\ 0 & 0 & 0 & 0 & 0 & 0 & 0 & 0 & 0 & 0 & 0 & 1 & 0 & 0 & 0 & 0 \\ 0 & 0 & 1 & 0 & 0 & 0 & 0 & 0 & 0 & 0 & 0 & 0 & 0 & 0 & 0 & 0 \\ 0 & 0 & 0 & 0 & 1 & 0 & 0 & 0 & 0 & 0 & 0 & 0 & 0 & 0 & 0 & 0 \\ 0 & 0 & 0 & 0 & 0 & 0 & 0 & 0 & 0 & 0 & 0 & 0 & 1 & 0 & 0 & 0 \\ 0 & 0 & 0 & 0 & 0 & 0 & 0 & 0 & 1 & 0 & 0 & 0 & 0 & 0 & 0 & 0 \\ 0 & 0 & 0 & 0 & 0 & 0 & 0 & 0 & 0 & 0 & 0 & 0 & 0 & 1 & 0 & 0 \\ 0 & 0 & 0 & 0 & 0 & 0 & 0 & 0 & 0 & 0 & 0 & 0 & 0 & 0 & 1 & 0 \\ 0 & 0 & 0 & 0 & 0 & 0 & 0 & 0 & 0 & 0 & 0 & 0 & 0 & 0 & 0 & 1 \end{bmatrix}$$

on $|W_{4,0}\rangle$. The states $|W_{4,5}\rangle$, $|W_{4,6}\rangle$ and $|W_{4,7}\rangle$ can be obtained by performing *ZIIZ*, *ZIZI* and *ZZII* operations on state $|W_{4,4}\rangle$, respectively. The state $|W_{4,8}\rangle$ can be obtained by performing *XXXX* operation on state $|W_{4,4}\rangle$. The state $|W_{4,9}\rangle$ can be obtained by performing *IIZZ* operation on state $|W_{4,8}\rangle$. The states $|W_{4,a}\rangle$ and $|W_{4,b}\rangle$ can be obtained by performing $-XXXX$ on states $|W_{4,6}\rangle$ and $|W_{4,5}\rangle$, respectively. All these $W_4$ states form a group of orthogonal bases in the 16-dimensional Hilbert space. Any 4-qubit state can be expressed as a linear combination of these 16 $W_4$ states.

## II. Details of W-state preparation based on entanglement swapping

Let the 4 Bell states prepared by Alice, Bob, Charlie and David are $|\phi^+\rangle_{AA'} = 1/\sqrt{2}(|00\rangle_{AA'} + |11\rangle_{AA'})$, $|\phi^+\rangle_{BB'} = 1/\sqrt{2}(|00\rangle_{BB'} + |11\rangle_{BB'})$, $|\phi^+\rangle_{CC'} = 1/\sqrt{2}(|00\rangle_{CC'} + |11\rangle_{CC'})$ and $|\phi^+\rangle_{DD'} = 1/\sqrt{2}(|00\rangle_{DD'} + |11\rangle_{DD'})$, respectively. The state of the system is

$$\begin{aligned}|\psi_S\rangle &= 1/4(|00\rangle_{AA'} + |11\rangle_{AA'})(|00\rangle_{BB'} + |11\rangle_{BB'})(|00\rangle_{CC'} + |11\rangle_{CC'})(|00\rangle_{DD'} + |11\rangle_{DD'}) \\ &= 1/4\{|0000\rangle_{ABCD}|0000\rangle_{A'B'C'D'} + |0001\rangle_{ABCD}|0001\rangle_{A'B'C'D'} + |0010\rangle_{ABCD}|0010\rangle_{A'B'C'D'} \\ &\quad + |0011\rangle_{ABCD}|0011\rangle_{A'B'C'D'} + |0100\rangle_{ABCD}|0100\rangle_{A'B'C'D'} + |0101\rangle_{ABCD}|0101\rangle_{A'B'C'D'} \\ &\quad + |0110\rangle_{ABCD}|0110\rangle_{A'B'C'D'} + |0111\rangle_{ABCD}|0111\rangle_{A'B'C'D'} + |1000\rangle_{ABCD}|1000\rangle_{A'B'C'D'} \\ &\quad + |1001\rangle_{ABCD}|1001\rangle_{A'B'C'D'} + |1010\rangle_{ABCD}|1010\rangle_{A'B'C'D'} + |1011\rangle_{ABCD}|1011\rangle_{A'B'C'D'} \\ &\quad + |1100\rangle_{ABCD}|1100\rangle_{A'B'C'D'} + |1101\rangle_{ABCD}|1101\rangle_{A'B'C'D'} + |1110\rangle_{ABCD}|1110\rangle_{A'B'C'D'} \\ &\quad + |1111\rangle_{ABCD}|1111\rangle_{A'B'C'D'}\} \end{aligned} \quad (S17)$$

The 16 basis states can be represented by 16 $W_4$ states as follows

$$|0001\rangle_{A'B'C'D'} = 1/2(|W_{4,0}\rangle + |W_{4,1}\rangle + |W_{4,2}\rangle + |W_{4,3}\rangle)_{A'B'C'D'} \quad (S18)$$

$$|0010\rangle_{A'B'C'D'} = 1/2(|W_{4,0}\rangle - |W_{4,1}\rangle - |W_{4,2}\rangle + |W_{4,3}\rangle)_{A'B'C'D'} \quad (S19)$$

$$|0100\rangle_{A'B'C'D'} = 1/2(|W_{4,0}\rangle - |W_{4,1}\rangle + |W_{4,2}\rangle - |W_{4,3}\rangle)_{A'B'C'D'} \quad (S20)$$

$$|1000\rangle_{A'B'C'D'} = 1/2(|W_{4,0}\rangle + |W_{4,1}\rangle - |W_{4,2}\rangle - |W_{4,3}\rangle)_{A'B'C'D'} \quad (S21)$$

$$|0000\rangle_{A'B'C'D'} = 1/2(|W_{4,4}\rangle + |W_{4,5}\rangle + |W_{4,6}\rangle + |W_{4,7}\rangle)_{A'B'C'D'} \quad (S22)$$

$$|1100\rangle_{A'B'C'D'} = 1/2(|W_{4,4}\rangle - |W_{4,5}\rangle - |W_{4,6}\rangle + |W_{4,7}\rangle)_{A'B'C'D'} \quad (S23)$$

$$|1010\rangle_{A'B'C'D'} = 1/2(|W_{4,4}\rangle - |W_{4,5}\rangle + |W_{4,6}\rangle - |W_{4,7}\rangle)_{A'B'C'D'} \quad (S24)$$

$$|1001\rangle_{A'B'C'D'} = 1/2(|W_{4,4}\rangle + |W_{4,5}\rangle - |W_{4,6}\rangle - |W_{4,7}\rangle)_{A'B'C'D'} \quad (S25)$$

$$|0011\rangle_{A'B'C'D'} = 1/2(|W_{4,8}\rangle + |W_{4,9}\rangle + |W_{4,a}\rangle + |W_{4,b}\rangle)_{A'B'C'D'} \quad (S26)$$

$$|0101\rangle_{A'B'C'D'} = 1/2(|W_{4,8}\rangle - |W_{4,9}\rangle - |W_{4,a}\rangle + |W_{4,b}\rangle)_{A'B'C'D'} \quad (S27)$$

$$|0110\rangle_{A'B'C'D'} = 1/2(|W_{4,8}\rangle - |W_{4,9}\rangle + |W_{4,a}\rangle - |W_{4,b}\rangle)_{A'B'C'D'} \quad (S28)$$

$$|1111\rangle_{A'B'C'D'} = 1/2(|W_{4,8}\rangle + |W_{4,9}\rangle - |W_{4,a}\rangle - |W_{4,b}\rangle)_{A'B'C'D'} \quad (S29)$$

$$|0111\rangle_{A'B'C'D'} = 1/2(|W_{4,c}\rangle + |W_{4,d}\rangle + |W_{4,e}\rangle + |W_{4,f}\rangle)_{A'B'C'D'} \quad (S30)$$

$$|1011\rangle_{A'B'C'D'} = 1/2(|W_{4,c}\rangle - |W_{4,d}\rangle - |W_{4,e}\rangle + |W_{4,f}\rangle)_{A'B'C'D'} \quad (S31)$$

$$|1101\rangle_{A'B'C'D'} = 1/2(|W_{4,c}\rangle - |W_{4,d}\rangle + |W_{4,e}\rangle - |W_{4,f}\rangle)_{A'B'C'D'} \quad (S32)$$

$$|1110\rangle_{A'B'C'D'} = 1/2(|W_{4,c}\rangle + |W_{4,d}\rangle - |W_{4,e}\rangle - |W_{4,f}\rangle)_{A'B'C'D'} \quad (S33)$$

Then,

$$\begin{aligned}|\psi_S\rangle = 1/4\{ & |W_{4,0}\rangle_{ABCD}|W_{4,0}\rangle_{A'B'C'D'} + |W_{4,1}\rangle_{ABCD}|W_{4,1}\rangle_{A'B'C'D'} + |W_{4,2}\rangle_{ABCD}|W_{4,2}\rangle_{A'B'C'D'} + \\ & |W_{4,3}\rangle_{ABCD}|W_{4,3}\rangle_{A'B'C'D'} + |W_{4,4}\rangle_{ABCD}|W_{4,4}\rangle_{A'B'C'D'} + |W_{4,5}\rangle_{ABCD}|W_{4,5}\rangle_{A'B'C'D'} + \\ & |W_{4,6}\rangle_{ABCD}|W_{4,6}\rangle_{A'B'C'D'} + |W_{4,7}\rangle_{ABCD}|W_{4,7}\rangle_{A'B'C'D'} + |W_{4,8}\rangle_{ABCD}|W_{4,8}\rangle_{A'B'C'D'} + \\ & |W_{4,9}\rangle_{ABCD}|W_{4,9}\rangle_{A'B'C'D'} + |W_{4,a}\rangle_{ABCD}|W_{4,a}\rangle_{A'B'C'D'} + |W_{4,b}\rangle_{ABCD}|W_{4,b}\rangle_{A'B'C'D'} + \\ & |W_{4,c}\rangle_{ABCD}|W_{4,c}\rangle_{A'B'C'D'} + |W_{4,d}\rangle_{ABCD}|W_{4,d}\rangle_{A'B'C'D'} + |W_{4,e}\rangle_{ABCD}|W_{4,e}\rangle_{A'B'C'D'} + \\ & |W_{4,f}\rangle_{ABCD}|W_{4,f}\rangle_{A'B'C'D'} \}\end{aligned} \quad (S34)$$

Therefore, from equation (S34) we obtain that when the state of particles $A', B', C'$ and $D'$ are projected into one of the $W_4$ states, the state of particles $A, B, C$ and $D$ will be in the corresponding $W_4$ state.

### III. Details of estimation of $Q_1$ and $e_1$

We compute the gains at different cases, respectively.
(1) Case 1: outputs deriving from four background counts

In this case, there is no successful photon transmission (with probability $(1-\eta_a)(1-\eta_b)(1-\eta_c)(1-\eta_d)$, where $\eta_i$ is the transmittance of participant $i$, $i = a,b,c,d$). Emma will give an output when there exist dark counts in four specific modes according to Table 1 (with probability $Y_0^4$) and there are no dark counts in other 12 modes (with probability $(1-Y_0)^{12}$). There are 12 (4) detection modes for states $|W_{4,0}\rangle$ and $|W_{4,c}\rangle$ ($|W_{4,1}\rangle$ and $|W_{4,d}\rangle$). Based on the post-selection scheme shown in Table 2, input states $0000, 0001, 0010$ and $0011$ (corresponding to the states $|W_{4,0}\rangle$ and $|W_{4,1}\rangle$), $1100, 1101, 1110$ and $1111$ (corresponding to $|W_{4,c}\rangle$ and $|W_{4,d}\rangle$) can be used to generate key. Each input state ($0000, 0001, ..., 1110$ or $1111$) is prepared with equal probability, i.e. each one with probability $1/16$. So the gain at this case is

$$1/16\{4(1-\eta_a)(1-\eta_b)(1-\eta_c)(1-\eta_d)\left[12Y_0^4(1-Y_0)^{12}+4Y_0^4(1-Y_0)^{12}\right]+$$
$$4(1-\eta_a)(1-\eta_b)(1-\eta_c)(1-\eta_d)\left[12Y_0^4(1-Y_0)^{12}+4Y_0^4(1-Y_0)^{12}\right]\} \tag{S35}$$
$$=8(1-\eta_a)(1-\eta_b)(1-\eta_c)(1-\eta_d)Y_0^4(1-Y_0)^{12}$$

While, qubit error will occur when input states are $0000, 0011, 1100$ and $1111$ with probability

$$4(1-\eta_a)(1-\eta_b)(1-\eta_c)(1-\eta_d)Y_0^4(1-Y_0)^{12}. \tag{S36}$$

(2) Case 2: outputs deriving from one photon count and three background counts

In this case, one count derives from a successful photon transmission and the other three counts derive from background. For an input photon, it can arrive at the SPD in different spatial and temporal modes, e.g. the state $a_{a,t_0}^\dagger|0\rangle$ of Alice's photon can be mapped into the state

$$a_{a,t_0}^\dagger|0\rangle \to 1/(4\sqrt{2})\left(-2a_{s,t_0}^\dagger - 2e^{i\delta}a_{s,t_1}^\dagger + 2ie^{i\delta}a_{u,t_1}^\dagger + 2ie^{i2\delta}a_{u,t_2}^\dagger - 2ia_{v,t_0}^\dagger \right.$$
$$\left. +2ie^{i\delta}a_{v,t_1}^\dagger - 2e^{i\delta}a_{w,t_1}^\dagger + 2e^{i2\delta}a_{w,t_2}^\dagger\right)|0\rangle, \tag{S37}$$

and the state $a_{a,t_1}^\dagger|0\rangle$ of Alice's photon can be mapped into the state

$$a_{a,t_1}^\dagger|0\rangle \to 1/(4\sqrt{2})\left(-2a_{s,t_1}^\dagger - 2e^{i\delta}a_{s,t_2}^\dagger + 2ie^{i\delta}a_{u,t_2}^\dagger + 2ie^{i2\delta}a_{u,t_3}^\dagger - 2ia_{v,t_1}^\dagger \right.$$
$$\left. +2ie^{i\delta}a_{v,t_2}^\dagger - 2e^{i\delta}a_{w,t_2}^\dagger + 2e^{i2\delta}a_{w,t_3}^\dagger\right)|0\rangle. \tag{S38}$$

When Alice's photon, e.g. in state $a_{a,t_0}^\dagger|0\rangle$, is detected and the input state is $|0000\rangle$, the output probability of Emma's device according to Table 1 is

$$\eta_a(1-\eta_b)(1-\eta_c)(1-\eta_d)\left(1/4\sqrt{2}\right)^2\left[2^2\times10+2^2\times3+2^2\times10+2^2\times3+\right.$$
$$\left. 2^2\times10+2^2\times3+2^2\times10+2^2\times3\right]Y_0^3(1-Y_0)^{12} \tag{S39}$$
$$=208/32\,\eta_a(1-\eta_b)(1-\eta_c)(1-\eta_d)Y_0^3(1-Y_0)^{12}$$

In equation (S39), 10 denotes the number of operators $a_{s,t_0}^\dagger, a_{u,t_1}^\dagger, a_{v,t_0}^\dagger a_{w,t_1}^\dagger$ in the detection modes of the W states $|W_{4,0}\rangle$ and $|W_{4,1}\rangle$; 3 denotes the number of operators $a_{s,t_1}^\dagger, a_{u,t_2}^\dagger, a_{v,t_1}^\dagger, a_{w,t_2}^\dagger$ in the detection modes of the W states $|W_{4,0}\rangle$ and $|W_{4,1}\rangle$. Similarly, we can obtain the output probability of other participants' photons' counts under different inputs state. So, the gain in this case is

$$1/16\{8\times208/32\left[\eta_a(1-\eta_b)(1-\eta_c)(1-\eta_d)+\eta_b(1-\eta_a)(1-\eta_c)(1-\eta_d)\right]+$$
$$(4\times208/32+4\times96/32)\left[\eta_c(1-\eta_a)(1-\eta_b)(1-\eta_d)+\eta_d(1-\eta_a)(1-\eta_b)(1-\eta_c)\right]\}Y_0^3(1-Y_0)^{12}$$
$$=1/16\{52\left[\eta_a(1-\eta_b)(1-\eta_c)(1-\eta_d)+\eta_b(1-\eta_a)(1-\eta_c)(1-\eta_d)\right]+ \tag{S40}$$
$$38\left[\eta_c(1-\eta_a)(1-\eta_b)(1-\eta_d)+\eta_d(1-\eta_a)(1-\eta_b)(1-\eta_c)\right]\}Y_0^3(1-Y_0)^{12}$$

In this case the error will occur with probability

$$1/16\{26[\eta_a(1-\eta_b)(1-\eta_c)(1-\eta_d)+\eta_b(1-\eta_a)(1-\eta_c)(1-\eta_d)]+ \\ 19[\eta_c(1-\eta_a)(1-\eta_b)(1-\eta_d)+\eta_d(1-\eta_a)(1-\eta_b)(1-\eta_c)]\}Y_0^3(1-Y_0)^{12} \tag{S41}$$

(3) Case 3: outputs deriving from two photon counts and two background counts

In this case, two counts derive from successful photon transmission and two counts derive from background. Any state of two input photons can evolve into a state superposed by different spatial and time modes, e.g. state $a_{a,t_0}^\dagger a_{b,t_0}^\dagger|0\rangle$ of Alice's and Bob's photons evolves into the state

$$a_{a,t_0}^\dagger a_{b,t_0}^\dagger|0\rangle \to 1/(2^2\cdot 2^2\cdot 2)\Big[-i4e^{i4\delta}(a_{u,t_2}^\dagger)^2 - 8e^{i4\delta}a_{u,t_2}^\dagger a_{w,t_2}^\dagger + i4e^{i4\delta}(a_{w,t_2}^\dagger)^2 + \\
8e^{i3\delta}a_{s,t_1}^\dagger a_{u,t_2}^\dagger - i8e^{i3\delta}a_{s,t_1}^\dagger a_{w,t_2}^\dagger - i8e^{i3\delta}a_{u,t_2}^\dagger a_{v,t_1}^\dagger - 8e^{i3\delta}a_{v,t_1}^\dagger a_{w,t_2}^\dagger + i4e^{i2\delta}(a_{s,t_1}^\dagger)^2 + \\
8e^{i2\delta}a_{s,t_1}^\dagger a_{v,t_1}^\dagger + i4e^{i2\delta}(a_{u,t_1}^\dagger)^2 - 8e^{i2\delta}a_{u,t_1}^\dagger a_{w,t_1}^\dagger - i4e^{i2\delta}(a_{v,t_1}^\dagger)^2 - i4e^{i2\delta}(a_{w,t_1}^\dagger)^2 - \\
8e^{i\delta}a_{s,t_0}^\dagger a_{u,t_1}^\dagger - i8e^{i\delta}a_{s,t_0}^\dagger a_{w,t_1}^\dagger - i8e^{i\delta}a_{u,t_1}^\dagger a_{v,t_0}^\dagger + 8e^{i\delta}a_{v,t_0}^\dagger a_{w,t_1}^\dagger - i4(a_{s,t_0}^\dagger)^2 + \\
8a_{s,t_0}^\dagger a_{v,t_0}^\dagger + i4(a_{v,t_0}^\dagger)^2\Big]|0\rangle \tag{S42}$$

From Table 1, we can obtain that there are 6, 6, 4, 4, 6 and 6 detection modes for states $a_{s,t_0}^\dagger a_{u,t_1}^\dagger, a_{s,t_0}^\dagger a_{v,t_0}^\dagger, a_{s,t_0}^\dagger a_{w,t_1}^\dagger, a_{u,t_1}^\dagger a_{v,t_0}^\dagger, a_{u,t_1}^\dagger a_{w,t_1}^\dagger$ and $a_{v,t_0}^\dagger a_{w,t_1}^\dagger$, respectively, corresponding to the states $|W_{4,0}\rangle$ and $|W_{4,1}\rangle$. Based on equation (S42) the output probability is

$$(1/32)^2 \eta_a\eta_b(1-\eta_c)(1-\eta_d)\times 64(6+6+4+4+6+6)Y_0^2(1-Y_0)^{12} \\
= 2\eta_a\eta_b(1-\eta_c)(1-\eta_d)Y_0^2(1-Y_0)^{12} \tag{S43}$$

Similarly, we can obtain the output probability of other participants' photons counts under different inputs state. So, the gain in this case is

$$1/16\{16\eta_a\eta_b(1-\eta_c)(1-\eta_d)+10[\eta_a\eta_c(1-\eta_b)(1-\eta_d)+\eta_b\eta_d(1-\eta_a)(1-\eta_c)]+ \\
9[\eta_a\eta_d(1-\eta_b)(1-\eta_c)+\eta_b\eta_c(1-\eta_a)(1-\eta_d)]+8\eta_c\eta_d(1-\eta_a)(1-\eta_b)\}Y_0^2(1-Y_0)^{12} \tag{S44}$$

In this case the error will occur with probability

$$1/32\{16\eta_a\eta_b(1-\eta_c)(1-\eta_d)+10[\eta_a\eta_c(1-\eta_b)(1-\eta_d)+\eta_b\eta_d(1-\eta_a)(1-\eta_c)]+ \\
9[\eta_a\eta_d(1-\eta_b)(1-\eta_c)+\eta_b\eta_c(1-\eta_a)(1-\eta_d)]+8\eta_c\eta_d(1-\eta_a)(1-\eta_b)\}Y_0^2(1-Y_0)^{12} \tag{S45}$$

(4) Case 4: outputs deriving from three photon counts and one background count

In this case, three counts derive from successful photon transmission and one count derives from background. Any state of three input photons can evolve into a state superposed by different spatial and time modes, e.g. the state $a_{a,t_0}^\dagger a_{b,t_0}^\dagger a_{c,t_0}^\dagger|0\rangle$ of Alice's, Bob's and Charlie's photons evolves into the state (partial terms are omitted for simplification)

$$a_{a,t_0}^\dagger a_{b,t_0}^\dagger a_{c,t_0}^\dagger |0\rangle \to 1/\left(2^3 \cdot 2^3 \cdot \sqrt{2^3}\right)\left[8e^{i6\delta}\left(a_{u,t_2}^\dagger\right)^3 - i8e^{i6\delta}\left(a_{u,t_2}^\dagger\right)^2 a_{w,t_2}^\dagger + 8e^{i6\delta}a_{u,t_2}^\dagger \left(a_{w,t_2}^\dagger\right)^2 - \right.$$
$$i8e^{i6\delta}\left(a_{w,t_2}^\dagger\right)^3 + i8e^{i5\delta}a_{s,t_1}^\dagger\left(a_{u,t_2}^\dagger\right)^2 - 16e^{i5\delta}a_{s,t_1}^\dagger a_{u,t_2}^\dagger a_{w,t_2}^\dagger + i24e^{i5\delta}a_{s,t_1}^\dagger\left(a_{w,t_2}^\dagger\right)^2 + \quad (S46)$$
$$\left....+i8e^{i\delta}\left(a_{v,t_0}^\dagger\right)^2 a_{w,t_1}^\dagger + i8\left(a_{s,t_0}^\dagger\right)^3 - 8\left(a_{s,t_0}^\dagger\right)^2 a_{v,t_0}^\dagger + i8a_{s,t_0}^\dagger\left(a_{v,t_0}^\dagger\right)^2 - 8\left(a_{v,t_0}^\dagger\right)^3\right]|0\rangle$$

In equation (S46), the states which may lead to successful outputs in Emma's device include: $a_{s,t_0}^\dagger a_{u,t_1}^\dagger a_{v,t_0}^\dagger$, $a_{s,t_0}^\dagger a_{u,t_1}^\dagger a_{v,t_1}^\dagger$, $a_{s,t_0}^\dagger a_{u,t_1}^\dagger a_{w,t_1}^\dagger$, $a_{s,t_0}^\dagger a_{u,t_1}^\dagger a_{w,t_2}^\dagger$, $a_{s,t_0}^\dagger a_{u,t_2}^\dagger a_{v,t_0}^\dagger$, $a_{s,t_1}^\dagger a_{u,t_1}^\dagger a_{w,t_1}^\dagger$, $a_{s,t_1}^\dagger a_{v,t_0}^\dagger a_{w,t_1}^\dagger$, $a_{u,t_1}^\dagger a_{v,t_1}^\dagger a_{w,t_1}^\dagger$, $a_{u,t_2}^\dagger a_{v,t_0}^\dagger a_{w,t_1}^\dagger$, $a_{s,t_0}^\dagger a_{v,t_0}^\dagger a_{w,t_1}^\dagger$, $a_{s,t_0}^\dagger a_{v,t_0}^\dagger a_{w,t_2}^\dagger$, $a_{u,t_1}^\dagger a_{v,t_0}^\dagger a_{w,t_1}^\dagger$, $a_{s,t_0}^\dagger a_{u,t_2}^\dagger a_{w,t_1}^\dagger$, $a_{s,t_0}^\dagger a_{v,t_1}^\dagger a_{w,t_1}^\dagger$, $a_{s,t_1}^\dagger a_{u,t_1}^\dagger a_{v,t_0}^\dagger$ and $a_{u,t_1}^\dagger a_{v,t_0}^\dagger a_{w,t_2}^\dagger$. Their coefficients are all $16/\left(2^3 \cdot 2^3 \cdot \sqrt{2^3}\right)$. In the former 12 terms, each one corresponds to 2 detection modes. In the later 4 terms, each one corresponds to 1 detection mode according to Table 1. When the input state is $a_{a,t_0}^\dagger a_{b,t_0}^\dagger a_{c,t_0}^\dagger |0\rangle$ the output probability of Emma's device is

$$\left(16/2^3 \cdot 2^3 \cdot \sqrt{2^3}\right)^2 \times (12 \times 2 + 4)\eta_a \eta_b \eta_c (1-\eta_d) Y_0 (1-Y_0)^{12} \quad (S47)$$

Similarly, we can obtain the output probabilities of other participants' photons counts under different inputs state. So, the gain in this case is

$$1/256\{34[\eta_a\eta_b\eta_c(1-\eta_d) + \eta_a\eta_b\eta_d(1-\eta_c)] + \\ 15[\eta_a\eta_c\eta_d(1-\eta_b) + \eta_b\eta_c\eta_d(1-\eta_a)]\}Y_0(1-Y_0)^{12} \quad (S48)$$

In this case the error will occur with probability

$$1/256\{17[\eta_a\eta_b\eta_c(1-\eta_d) + \eta_a\eta_b\eta_d(1-\eta_c)] + \\ 7[\eta_a\eta_c\eta_d(1-\eta_b) + \eta_b\eta_c\eta_d(1-\eta_a)]\}Y_0(1-Y_0)^{12} \quad (S49)$$

(5) Case 5: outputs deriving from four photon counts

In this case, there exist successful outputs when the input state is one of $|0010\rangle, |0001\rangle, |1101\rangle$ or $|1110\rangle$. State $|0001\rangle$ can be projected into states $|W_{4,0}\rangle$ or $|W_{4,1}\rangle$ with probability $1/4$. Let the identification probabilities of $|W_{4,0}\rangle$ and $|W_{4,1}\rangle$ be $D_{p0}$ and $D_{p1}$, respectively. When input state is $|0001\rangle$ the output probability is

$$1/4\,\eta_a\eta_b\eta_c\eta_d(D_{p0} + D_{p1})(1-Y_0)^{12} \quad (S50)$$

So, the output probability in the case is

$$1/16 \times 4 \times 1/4\,\eta_a\eta_b\eta_c\eta_d(D_{p0} + D_{p1})(1-Y_0)^{12} \\ = 1/16\,\eta_a\eta_b\eta_c\eta_d(D_{p0} + D_{p1})(1-Y_0)^{12} \quad (S51)$$

There is no error in this case.
By adding all output probabilities in five cases together, the gain is

$$\begin{aligned}
Q_1 = & 8(1-\eta_a)(1-\eta_b)(1-\eta_c)(1-\eta_d)Y_0^4(1-Y_0)^{12} + \\
& 1/16\{52[\eta_a(1-\eta_b)(1-\eta_c)(1-\eta_d)+\eta_b(1-\eta_a)(1-\eta_c)(1-\eta_d)] + \\
& 38[\eta_c(1-\eta_a)(1-\eta_b)(1-\eta_d)+\eta_d(1-\eta_a)(1-\eta_b)(1-\eta_c)]\}Y_0^3(1-Y_0)^{12} + \\
& 1/16\{16\eta_a\eta_b(1-\eta_c)(1-\eta_d)+10[\eta_a\eta_c(1-\eta_b)(1-\eta_d)+\eta_b\eta_d(1-\eta_a)(1-\eta_c)] + \\
& 9[\eta_a\eta_d(1-\eta_b)(1-\eta_c)+\eta_b\eta_c(1-\eta_a)(1-\eta_d)]+8\eta_c\eta_d(1-\eta_a)(1-\eta_b)\}Y_0^2(1-Y_0)^{12} + \\
& 1/256\{34[\eta_a\eta_b\eta_c(1-\eta_d)+\eta_a\eta_b\eta_d(1-\eta_c)] + \\
& 15[\eta_a\eta_c\eta_d(1-\eta_b)+\eta_b\eta_c\eta_d(1-\eta_a)]\}Y_0(1-Y_0)^{12} + \\
& 1/16\eta_a\eta_b\eta_c\eta_d(D_{p0}+D_{p1})(1-Y_0)^{12}
\end{aligned} \quad (S52)$$

, and the QBER is

$$\begin{aligned}
e_1 = & 4(1-\eta_a)(1-\eta_b)(1-\eta_c)(1-\eta_d)Y_0^4(1-Y_0)^{12} + \\
& 1/16\{26[\eta_a(1-\eta_b)(1-\eta_c)(1-\eta_d)+\eta_b(1-\eta_a)(1-\eta_c)(1-\eta_d)] + \\
& 19[\eta_c(1-\eta_a)(1-\eta_b)(1-\eta_d)+\eta_d(1-\eta_a)(1-\eta_b)(1-\eta_c)]\}Y_0^3(1-Y_0)^{12} + \\
& 1/32\{16\eta_a\eta_b(1-\eta_c)(1-\eta_d)+10[\eta_a\eta_c(1-\eta_b)(1-\eta_d)+\eta_b\eta_d(1-\eta_a)(1-\eta_c)] + \\
& 9[\eta_a\eta_d(1-\eta_b)(1-\eta_c)+\eta_b\eta_c(1-\eta_a)(1-\eta_d)]+8\eta_c\eta_d(1-\eta_a)(1-\eta_b)\}Y_0^2(1-Y_0)^{12} + \\
& 1/256\{17[\eta_a\eta_b\eta_c(1-\eta_d)+\eta_a\eta_b\eta_d(1-\eta_c)] + \\
& 7[\eta_a\eta_c\eta_d(1-\eta_b)+\eta_b\eta_c\eta_d(1-\eta_a)]\}Y_0(1-Y_0)^{12}
\end{aligned} \quad (S53)$$

We assume that four optical fiber links and four detectors are identical, i.e. $\eta_a = \eta_c = \eta_d = \eta_b = \eta$. So, equations (S52) and (S53) can be simplified as

$$\begin{aligned}
Q_1 = & (1-Y_0)^{12}\left[1024(1-\eta)^4 Y_0^4 + 1440\eta(1-\eta)^3 Y_0^3 + 496\eta^2(1-\eta)^2 Y_0^2 + \right. \\
& \left. 49\eta^3(1-\eta)Y_0 + 8(D_{p0}+D_{p1})\eta^4\right]/128
\end{aligned} \quad (S54)$$

$$\begin{aligned}
e_1 = & \frac{1}{16Q_1}(1-Y_0)^{12}\left[64(1-\eta)^4 Y_0^4 + 90\eta(1-\eta)^3 Y_0^3 + 31\eta^2(1-\eta)^2 Y_0^2 + \right. \\
& \left. 3\eta^3(1-\eta)Y_0\right]
\end{aligned} \quad (S55)$$

1. Verstraete, F., Dehaene, J., De Moor, B. & Verschelde, H. Four qubits can be entangled in nine different ways. *Phys. Rev. A* **65,** 052112 (2002).